# Deciphering seasonal depression variations and interplays between weather changes, physical activity, and depression severity in real-world settings: Learnings from RADAR-MDD longitudinal mobile health study

Yuezhou Zhang[1], Amos A. Folarin[1,2,3,4,5], Yatharth Ranjan[1], Nicholas Cummins[1], Zulqarnain Rashid[1], Pauline Conde[1], Callum Stewart[1], Shaoxiong Sun[1,6], Srinivasan Vairavan[7], Faith Matcham[8,9], Carolin Oetzmann[8], Sara Siddi[10,11], Femke Lamers[12,13], Sara Simblett[14], Til Wykes[14,15], David C. Mohr[16], Josep Maria Haro[10,11], Brenda W.J.H. Penninx[12,13], Vaibhav A. Narayan[7], Matthew Hotopf[3,8,15], Richard J.B. Dobson[1,2,3,4,5], Abhishek Pratap[1,17,18], RADAR-CNS consortium[19]

[1]Department of Biostatistics & Health Informatics, Institute of Psychiatry, Psychology and Neuroscience, King's College London, London, United Kingdom
[2]Institute of Health Informatics, University College London, London, United Kingdom
[3]NIHR Biomedical Research Centre at South London and Maudsley, NHS Foundation Trust, London, United Kingdom
[4]Health Data Research UK London, University College London, London, United Kingdom
[5]NIHR Biomedical Research Centre at University College London Hospitals, NHS Foundation Trust, London, United Kingdom
[6]Department of Computer Science, University of Sheffield, Sheffield, United Kingdom
[7]Janssen Research and Development LLC, Titusville, NJ, United States
[8]Department of Psychological Medicine, Institute of Psychiatry, Psychology and Neuroscience, King's College London, London, United Kingdom
[9]School of Psychology, University of Sussex, Falmer, United Kingdom
[10]Teaching Research and Innovation Unit, Parc Sanitari Sant Joan de Déu, Fundació Sant Joan de Déu, Barcelona, Spain
[11]Centro de Investigación Biomédica en Red de Salud Mental, Madrid, Spain
[12]Department of Psychiatry, Amsterdam UMC, Vrije Universiteit, Amsterdam, Netherlands
[13]Mental Health Program, Amsterdam Public Health Research Institute, Amsterdam, Netherlands
[14]Department of Psychology, Institute of Psychiatry, Psychology and Neuroscience, King's College London, London, United Kingdom
[15]South London and Maudsley NHS Foundation Trust, London, United Kingdom
[16]Center for Behavioral Intervention Technologies, Department of Preventive Medicine, Northwestern University, Chicago, IL, United States
[17]Boehringer Ingelheim Pharmaceuticals, Inc., Ridgefield, CT, United States
[18]University of Washington, Seattle, United States
[19]www.radar-cns.org

Corresponding authors: Richard J.B. Dobson (richard.j.dobson@kcl.ac.uk) and Abhishek Pratap (abhishek.vit@gmail.com)

**Abstract**

Prior research has shown that changes in seasons and weather can have a significant impact on depression severity. However, findings are inconsistent across populations, and the interplay between weather, behavior, and depression has not been fully quantified. This study analyzed real-world data from 428 participants (a subset; 68.7% of the cohort) in the RADAR-MDD longitudinal mobile health study to investigate seasonal variations in depression (measured through a remote validated assessment - PHQ-8) and examine the potential interplay between dynamic weather changes, physical activity (monitored via wearables), and depression severity. The clustering of PHQ-8 scores identified four distinct seasonal variations in depression severity: one stable trend and three varying patterns where depression peaks in different seasons. Among these patterns, participants within the stable trend had the oldest average age (p=0.002) and the lowest baseline PHQ-8 score (p=0.003). Mediation analysis assessing the indirect effect of weather on physical activity and depression showed significant differences among participants with different affective responses to weather. Specifically, the temperature and day length significantly influenced depression severity, which in turn impacted physical activity levels (p<0.001). For instance, among participants with a negative correlation between depression severity and temperature, a 10 °C increase led to a total daily step count rise of 655.4, comprised of 461.7 steps directly due to the temperature itself and 193.7 steps because of decreased depressive severity (1.9 decrease in PHQ-8). In contrast, for those with a positive correlation, a 10°C rise directly led to a 262.3-step rise; however, it was offset by a 141.3-step decrease due to increased depression severity (2.1 increase in PHQ-8) from higher temperatures, culminating in an insignificant overall increase of 121 steps. These findings illustrate the heterogeneity in individuals' seasonal depression variations and responses to weather, underscoring the necessity for personalized approaches to help understand the impact of environmental factors on the real-world effectiveness of behavioral treatments.

# Introduction

Depression, recognized as the most prevalent mental disorder worldwide, is a leading cause of disability [1]. Despite its substantial economic and social burden [2], the underlying etiology, pathophysiology, and effective treatments remain partially understood [3-5]. Previous research has shown that the variability of environmental factors (e.g., seasons and weather conditions) [6-8] and individual behaviors (e.g., physical activity, social interaction, and sleep) [9-13] can impact the severity of depression symptoms. However, the impact of such external environmental factors on depression may manifest differently across individuals [14-17]. Furthermore, there could be potential interplays between environmental factors, behaviors, and depression severity (e.g., weather may affect depression status, which in turn impacts physical activity) that are not fully understood so far. Further research is needed to understand how the short- and long-term changes in weather and real-world behaviors are interconnected with depression symptom severity and may impact individuals differently.

Previous studies have shown the significant impact of weather variations and seasonality on individual depression severity [6-8, 18-20]. Seasonal Affective Disorder (SAD) is conceptualized as a subtype of depression characterized by different symptomatology and marked seasonal patterns of recurrent depressive episodes [18, 19]. However, the literature shows notable inconsistencies in seasonal variation at the population level. While some studies report increased depression symptoms in winter [19, 21, 22], others demonstrated peaks in depression severity during spring [20, 23, 24], summer [20, 25] and autumn [23], with still others failing to find any significant seasonal effects [26, 27]. Regarding weather-related factors, some studies linked severe depressive symptoms to higher rainfall, lower temperature, reduced sunlight, and overcast conditions [20, 28]. In contrast, some studies have found contradictory findings [29, 30] or no significant correlations [31-33]. The heterogeneity of individual affective responses to varying weather conditions within study cohorts might be one of the potential reasons for these inconsistent findings [32]. For example, if a subgroup of individuals responds positively to changes in weather or season (e.g., temperature) while another subgroup responds negatively or shows no response, the overall cohort level effect may be non-significant [32]. Furthermore, even

SAD diagnoses are differentiated between winter and summer SAD [34, 35], indicating individual differences in responses to daylight exposure [32]. Therefore, it is essential to investigate different seasonal variation patterns of depression and different subtypes of responses to weather changes within populations.

Varying weather patterns can also significantly impact real-world behaviors, especially physical activity [36, 37]. For example, research has shown a positive correlation between physical activity levels and both ambient temperature and day length, while a negative correlation exists with precipitation and wind speed [37-39]. Physical activity and depression are negatively associated with a bidirectional manner. Depression may lead to decreased levels of physical activity, while reduced physical activity is also a known risk factor for depression [10, 40-42]. Consequently, the impact of weather on depression severity and physical activity may be interlinked (see Supplementary Figure 1 for a schematic diagram). For instance, weather conditions may directly affect physical activity levels or indirectly influence them by lowering mood. Such interplays between weather and physical activity on behavior can be systematically explored with real-world long-term observational data using mediation analysis. The analysis provides insights into how one variable can affect another through the intermediary role of a mediator [43]. A comprehensive understanding of how weather affects both depression and physical activity could refine personalized prevention strategies for depression, such as tailored use of exercise [42].

Despite the importance of these interconnected relationships, to our knowledge, no studies have explored the mediating effects between depression, physical activity, and environmental factors such as weather. Furthermore, most previous research on weather and depression has been cross-sectional, focusing on broad cohort associations [6, 20, 28, 30], which limits understanding of individual differences and within-individual associations [31]. Additionally, many previous studies have relied on participants recalling their emotional and behavioral states over months or years, which may introduce the subjective recall bias [21, 44, 45]. Longitudinal observational studies that use mobile technologies offer a cost-efficient method to monitor participants' behaviors, health status, and environmental variables over a longer term [46, 47].

To address these research gaps, this study aims to investigate the seasonal variations of depression and the interplay between dynamic weather changes, physical activity, and depression severity. This study leveraged data from a large longitudinal mobile health study, Remote Assessment of Disease and Relapse Major Depressive Disorder (RADAR-MDD), utilizing up to two years of individual data [48]. Specifically, we investigated the following research questions (RQ):

RQ1: How does depression severity vary across seasons?

RQ2: What are the mediating effects between weather conditions, physical activity, and depression severity?

RQ3: Are there subgroups of participants whose depression severity is differentially impacted by weather changes, and if these subgroups exhibit varying mediating effects on physical activity and depression outcomes?

## Methods

### Participants and Settings

This study leveraged data from the RADAR-MDD research program, which explored the effectiveness of remote mobile technologies for monitoring depression and predicting relapse in MDD [48]. The RADAR-MDD study recruited 623 participants from three study sites in the United Kingdom, Spain, and the Netherlands and followed them for up to 2 years [49]. Recruitment spanned November 2017 to June 2020, with data collection concluding in April 2021 [49]. Due to rolling enrollment, the follow-up duration varied from 11 months to 24 months [49]. Utilizing the RADAR-base open-source platform, the RADAR-MDD program concurrently gathered both active (e.g., questionnaires) and passive (e.g., smartphone and Fitbit device) data [50].

The RADAR-MDD protocol was co-developed with a patient advisory board (PAB), who shared their opinions on several aspects of the study, including the choice and frequency of survey measures, the usability of the study app, participant-facing documents, selection of optimal participation incentives, and the deployment of wearable device as well as the data analysis plan.

Ethical approvals were obtained from the Camberwell St. Giles Research Ethics Committee (17/LO/1154) in the UK, the Fundacio Sant Joan de Deu Clinical Research Ethics Committee (CI: PIC-128-17) in Spain, and the Medische Ethische Toetsingscommissie VUmc (2018.012–NL63557.029.17) in the Netherlands.

**Measures**

Depression severity: The severity of participants' depression symptoms was measured biweekly via smartphones using the 8-item Patient Health Questionnaire (PHQ-8) [51]. The PHQ-8 contains eight questions, with the total score ranging from 0 to 24, indicating increasing severity [51].

Seasons: The seasons were divided based on the meteorological season calendar: spring begins on March 1, summer on June 1, autumn on September 1, and winter on December 1 [30]. Each PHQ-8 questionnaire is assigned a season based on its completion date.

Weather variables: Throughout the RADAR-MDD study period, the city information relating to the participant's location was captured at the time of PHQ-8 assessment completion via the study app [50]. Utilizing coarse geospatial information, individual-level weather data were retrospectively sourced from the OpenWeather historical API (https://openweathermap.org/api/). We extracted the daily averages for several key meteorological parameters—ambient temperature (in Celsius), atmospheric pressure (in hPa), humidity (%), wind speed (in meters/sec), cloudiness (%), and day length (the time between sunrise and sunset, measured in hours)—specific to the dates on which the PHQ-8 questionnaires were completed.

Physical Activity: As part of the RADAR-MDD study, participants were instructed to wear a Fitbit wristband to monitor their daily behaviors [48]. As the PHQ-8 is designed to evaluate depression severity over the past two weeks [51], we calculated the average daily step count in a two-week window prior to the PHQ-8 to approximate participants' physical activity levels. This approach linked the participant's physical activity level with both the depression assessment and weather conditions.

Covariates: Since participants' depression severity and physical activity levels can be affected by several socio-demographic factors [52, 53], we considered age, gender, years in education, having children, employment status, marital status, annual income level, and study site as covariates in subsequent mediation analysis. Furthermore, as the data collection was conducted during the COVID-19 pandemic, we also considered a binary covariate to represent whether there is a national lockdown to partially reduce the effects of COVID-19 restrictions [54].

**Data Analysis**

**Data Inclusion Criteria**

To ensure data integrity, we filtered out incomplete submissions of the PHQ-8 questionnaire. Since this study included analyzing seasonal variations, participants were excluded from the study if they failed to complete the PHQ-8 questionnaire during any given season.

**Clustering Analysis—Seasonal Variations in Depression Severity (RQ1)**

We calculated the average PHQ-8 scores for each participant in every season and then mean-centered these seasonal averages by subtracting the participant's overall mean PHQ-8 score. To identify distinct seasonal patterns within the cohort, we employed the K-means clustering method, with the optimal number of clusters determined by the elbow method [55]. After clustering, we conducted a comparative analysis of participants' demographics and PHQ-8 scores across the identified clusters using the Kruskal-Wallis (KW) test [56]. Additionally, for each identified cluster, we reported the seasonal changes in physical activity levels (measured in daily step count). Furthermore, the clustering analysis was also separately performed on each study site.

**Multilevel Mediation Analysis—Exploring Interplays between Weather, Physical Activity, and Depression (RQ2)**

Mediation analysis is a statistical approach used for understanding the mechanisms underlying the exposure-outcome relationship by partitioning the total effect into direct and indirect components through an intervening variable, commonly referred to as a mediator [57]. Since our data are longitudinal (each participant has multiple measurements), we employed the 1-1-1 multilevel mediation model, in which all measures were collected at the individual level [58]. This model estimates the following effects: (i) total effect—the exposure-outcome effect without considering the effect of the mediator (c path), (ii) direct effect—the effect of the exposure on the outcome when adjusted for the mediator (c' path), (iii) indirect effect—the effect of the exposure on the outcome through the mediator (a*b), (iv) the effect of exposure on the mediator (a path), and (v) the effect of the mediator on the outcome (b path) (Figure 1).

We defined each weather measure as an exposure variable and constructed two distinct mediation models for each weather variable. The first model (Model 1) considered depression symptom severity (PHQ-8) as the mediator and physical activity (Step Count) as the outcome, while the second (Model 2) treated physical activity as the mediator and depression severity as the outcome variable. The covariates mentioned above were included to mitigate the risk of biased effect estimations [57].

**Subgroup Analysis for Distinct Affective Responses to Varying Weather Conditions (RQ3)**

We hypothesized that the presence of different affective responses to varying weather conditions within a population may obscure the correlation between weather conditions and depression at the cohort level. Similarly, mediating effects that may be significant within specific weather response subgroups could become obscured or attenuated when analyzed across the entire cohort. To identify distinct subgroups, we calculated the Spearman correlation coefficients between each participant's PHQ-8 scores and each weather variable. Following the guidelines for interpreting correlation coefficients [59], we set ±0.2 as the threshold for weak correlation, which is suitable for our current exploratory analysis with the limited sample size to sensitively detect potential associations warranting future in-depth investigation. Accordingly, participants were categorized into 'Positive Response' (Spearman Coefficient > 0.2), 'Negative Response' (Spearman Coefficient < -0.2), and 'Unaffected' (-0.2 ≤ Spearman Coefficient ≤ 0.2) subgroups, indicating positive, negative, and no significant affective responses to weather conditions, respectively. Subsequently, the mediation model was applied independently to the three subgroups for each weather variable.

# Results

**Data Overview**

We analyzed 12,490 PHQ-8 questionnaires with corresponding weather information and Fitbit step count recordings from 428 participants per the data inclusion criteria (see Methods). The selected cohort had a median age of 50.0 years [IQR: 32.0, 60.0] and was predominantly female (N=331, 77.3%), with a median PHQ-8 score of 9.5 [IQR: 6.0, 13.7]. Participants' socio-demographics (except for gender) and PHQ-8 score distributions were significantly different across the three study sites. Participants from CIBER (Spain) were the oldest and had the highest PHQ-8 scores. Further detailed socio-demographic information is provided in Table 1. Additionally, a sensitivity analysis showed no significant socio-demographic differences between the selected cohort in this study and the full RADAR-MDD study cohort. See supplementary Table 1 for additional details.

Figure 2 illustrates the seasonal fluctuations in weather variables across the three countries. The data reveal distinct seasonal patterns in ambient temperature and day length, peaking in summer and reaching the lowest values in winter. Spain exhibits higher temperatures than the UK and the Netherlands. The UK and the Netherlands experienced higher humidity and cloudiness during autumn and winter, compared to spring and summer. Conversely, Spain experienced minimal seasonal variations in humidity and cloudiness, with slightly lower values in summer and generally lower levels compared to the other countries. Furthermore, atmospheric pressure and wind speed are relatively stable across seasons, with minor variations.

**Distinct Seasonal Patterns in Depression Severity Variation**

The unsupervised clustering of mean-centered PHQ-8 data revealed four distinct seasonal variations in depression severity. Figure 3 illustrates the seasonal changes in mean-centered PHQ-8 scores for participants in Clusters 1-4, as well as corresponding changes in physical activity levels (measured in daily steps). The centered PHQ-8 scores represent deviations from each participant's mean PHQ-8 score. In Cluster 1 (N=199), participants exhibited minimal seasonal variation in their PHQ-8 scores, with only an average difference of 0.99 points between the highest and lowest scores across all seasons. Cluster 2 (N=93) displayed the highest PHQ-8 scores in spring (1.89 points above the mean), and the lowest in autumn (1.73 points below the mean). For Cluster 3 (N=73), participants experienced their peak PHQ-8 scores in winter (2.21 points above the mean), and the lowest in summer (2.13 points below the mean). Lastly, participants in Cluster 4 (N=63) showed the highest PHQ-8 scores in autumn (2.64 points above the mean), and the lowest in winter (1.43 points below the mean).

In addition, participants' characteristics were significantly different across the four identified clusters. Participants in Cluster 1 were the oldest (median age: 54.0 years [IQR: 37.5, 62.0]), while those in Cluster 3 were the youngest (median age: 41.0 years [IQR: 30.0, 57.0]) (KW test: p = 0.002) (Figure 3e). The baseline PHQ-8 was lowest in Cluster 1 (median: 9.0 [6.0,13.0]), and highest in Cluster 2 (median: 13.0 [8.0,16.0]) (KW test: p = 0.002) (Figure 3f). The proportion of female participants was lower in Cluster 1 (70.9%) compared to other clusters (Cluster 2: 83.9%, Cluster 3: 86.3%, and Cluster 4: 77.8%) (KW test: p = 0.02). Additionally, the proportion of participants from each study site significantly varied across the clusters (KW test: p = 0.02). Details of the comparative analysis across the clusters are provided in Supplementary Table 2.

Additionally, in Cluster 1, where the seasonal PHQ-8 variation is the smallest, the physical activity pattern appears to follow a seasonal trend, peaking in summer and declining to its lowest point in winter.

Despite differences in clustering results across sites, each site showed distinct seasonal patterns of depression severity, including a stable seasonal pattern characterized by minimal fluctuations in PHQ-8 scores across seasons. Supplementary Figures 2-4 show the sensitivity of the clustering analyses across three study sites.

**Mediating Effects of Weather Conditions on Physical Activity and Depression Severity**

While the indirect effects of weather changes on physical activity via depression severity (Model 1) were modest in the entire study cohort, these impacts, especially for temperature and day length, became relatively substantial in subgroups (Table 2). Specifically, to test the mediating effects among participants with different affective responses to weather conditions, participants were assigned to three subgroups based on individual Spearman coefficients between depression severity (PHQ-8) and each weather variable (see Methods). Temperature and day length significantly affected depression severity, which in turn influenced the physical activity in the Positive and Negative Response subgroups, and these effects were opposite (Figure 4 and Table 2). Detailed results of these two models are reported below.

Temperature: For each 10 °C increase, divergent effects on PHQ-8 scores were observed across subgroups: a 2.1-point increase in the Positive Response subgroup (p < 0.001), a 1.9-point decrease in the Negative Response subgroup (p < 0.001), and stability in the Unaffected subgroup. Due to different responses to temperature variations (a path) and the inverse relationship between depression severity and physical activity (b path), a 10°C increase in temperature led to opposite indirect effects on daily step counts through depression severity: a decrease of 141.3 steps in the Positive Response subgroup and an increase of 193.7 steps in the Negative Response subgroup (p < 0.001), and modest effects (0.6 steps; p=0.94) in Unaffected subgroup. The direct effects of temperature on physical activity were significant and positive across all subgroups (increments of 262.3 steps (p=0.02), 461.7 steps (p<0.001), and 471.0 steps (p<0.001) per 10 °C rise in Positive, Negative Response, and Unaffected subgroups, respectively). Thus, the total effect, integrating direct and indirect influences, showed an increase of 262.3-141.3 = 121.0 steps in the Positive Response subgroup (not significant), 461.7+193.7 = 655.4 steps in the Negative Response subgroup (p < 0.001), and 471.0 - 0.6 = 470.4 steps in the Unaffected subgroup (p < 0.001) per 10 °C rise.

Day length: Observations across subgroups revealed that each additional hour of day length resulted in varying PHQ-8 score changes: a 0.39-point increase in the Positive Response subgroup (p < 0.001), a 0.42-point decrease in the Negative Response subgroup (p < 0.001), and insignificant changes in the Unaffected subgroup. Similarly, for the indirect effects, an extra hour of day length contributed to contrasting daily step count outcomes: a reduction of 42.47 steps in the Positive Response subgroup and an increment of 36.82 steps in the Negative Response subgroup (p < 0.001). The direct influence of day length on physical activity was also uniformly positive and significant across subgroups (each hour increment led to increases of 64.61 (p=0.008), 94.85 (p<0.001), and 43.53 (p<0.001) daily steps in Positive, Negative Response, and

Unaffected subgroups, respectively). Consequently, the total effect demonstrated step count increment of 22.14 steps (not significant) in the Positive Response subgroup, 131.67 steps ($p < 0.001$) in the Negative Response subgroup, and 44.08 steps ($p < 0.001$) in the Unaffected subgroup per additional hour of day length.

Furthermore, for other weather variables such as humidity, cloudiness, pressure, and wind speed, the mediating effects on physical activity and their impacts on depression severity were relatively modest (Table 2), considering the annual variability range of these variables (Figure 2). On the other hand, the examination of how weather conditions affect depression severity through physical activity (Model 2) found the indirect effects were modest, indicating the impact of weather on depression is predominantly direct. Detailed results of Model 2 are available in Supplementary Table 3.

## Discussion

Analyzing data from a large longitudinal mobile health study collected in real-world settings, this present study identified distinct seasonal patterns of depression severity variations within the study cohort. We further explored the interplay between weather, physical activity, and depression severity across the entire cohort and among subgroups defined by different affective responses to weather. Our findings reveal that both temperature and day length significantly influenced depression severity, which in turn affected physical activity levels. Intriguingly, these indirect influences manifest differently or even oppositely across subgroups of participants. These findings not only enhance our understanding of the mechanisms of how weather affects depression severity and physical activity but also support the existence of different seasonal variations in depression severity and diverse responses to weather within a population. One of the primary strengths of our study is the utilization of longitudinal data collected using remote mobile technologies that captured variations in weather, behaviors, and depression severity at the individual level and minimized the recall bias.

This study offers valuable insights for both clinicians and data scientists. Past clinical studies revealed that preventive interventions such as antidepressants or light therapy before the season wherein symptoms usually emerge can prevent SAD [60, 61]. Additionally, regular exercise has been reported to prevent depression [42]. However, the effectiveness of these interventions varied across individuals [62-64], indicating preventative treatment must individualize treatment choices [65]. Our findings suggest that understanding distinct seasonal variations in depression and the impact of weather may refine the personalized prevention strategies for depression and understand the individual differences in the effectiveness of depression treatments. From the data analysis perspective, our data-driven results suggest that incorporating weather variables into future depression models is essential. Additionally, due to observed variations in weather conditions across different regions (Figure 2) and their potential distinct impacts on depression, it is necessary to consider external environmental factors such as weather when aggregating real-world data across studies, sites, and geospatial regions. Furthermore, the heterogeneity of individual affective responses to weather highlights the critical need for the subgroup or personalized analysis to enhance the effectiveness of models in future mental health research.

We identified distinct seasonal patterns of depression severity variation within our cohort via clustering (Figure 3). This observation supports the existence of multiple seasonal patterns within a population, potentially explaining the reason for inconsistent findings regarding seasonality's effects on depression reported in prior research [20-22, 27, 66]. Within these identified patterns, both across the entire cohort and at individual study sites, we pinpointed a cluster of participants characterized by minimal seasonal variations in their depression severity, presenting a relatively stable condition of depression throughout the year (Cluster 1 in Figure 3 and Supplementary Figures 2-4). This finding indicates that seasonality's influence on depression may be limited to a subset of individuals, potentially explaining the small effect size or lack of significant correlation between seasonality and depression reported in the general populations [22, 26, 27]. Comparing participants' characteristics across these seasonal patterns revealed that participants with minimal seasonal variations in depression severity were typically older. This aligns with findings from a meta-analysis of 20 population studies, which suggested that SAD is more prevalent among young adults [19]. Notably, the proportion of females was higher in clusters with larger seasonal variation (Clusters 2 and 3 in Figure 3), echoing previous research indicating that females showed stronger seasonal variation in depression than males [19, 22, 67]. While high-dimensional seasonal analyses shed light on the distinct seasonal variations in depression, variations in depression status may be influenced by multiple factors, such as socio-demographics, COVID-19, and personal issues. To uncover external environmental factors more closely associated with depression, we employed multilevel mediation analysis, using weather data of finer granularity and taking covariates into consideration.

Changes in temperature and day length were seen to notably impact participants' depression severity differentially with significant indirect influence on physical activity levels. The overall impact of weather on physical activity can be decomposed into direct impacts caused by the weather itself and indirect impacts resulting from changes in depression severity due to weather conditions. Specifically, for participants who experience reduced depression severity with a rise in temperatures (Negative Response subgroup), the direct impact of temperature and its indirect effects through depression were consistent, leading to a more pronounced overall effect on physical activity. In contrast, within the Positive Response subgroup, a rise in temperatures led to increasing depression severity. However, the direct influence of temperature on physical activity was not consistent with the indirect impact of change in depression severity. These contrasting direct and indirect effects led to the overall combined effect being insignificant on physical activity levels. Similar findings were observed with the day length variable. These significant indirect influences illustrate the mediating role of depression severity in the relationship between weather conditions and physical activity levels. Moreover, the differential impacts observed across subgroups underscore the heterogeneity of individual responses to weather changes. Prior studies have reported inconsistent affective weather responses for temperature and day length. Klimstra et al found that 16.8% of participants experienced improved mood with higher temperatures and more sunshine, while 26.8% showed the opposite effect, with the remaining 47.8% unaffected, in a cohort of 497 participants over 30 days of survey and weather data [32]. Moreover, some studies have shown that increased temperature can alleviate depression symptoms [20], while others suggest high temperatures heighten the risk of mental disorders-related admissions and suicide [68]. Regarding day length, certain studies have identified an improvement in depression symptoms with increased daylight [69], supporting the effectiveness of light therapy in treating depression [63]. Conversely, other research has reported poorer sleep and mood in some individuals during summer, potentially linked to summer SAD [70, 71].

Beyond the indirect effects, weather variations also directly and significantly affect physical activity. The estimations of direct effects in mediation analysis revealed that elevated temperatures and extended day length were also associated with increased physical activity levels, which aligns with previous studies [37-39]. However, the magnitude of the direct effects of weather conditions on physical activity varied across subgroups, suggesting that individuals' preferences for certain weather conditions might influence how these conditions affect their behaviors. Furthermore, we observed that higher levels of depression severity were correlated with diminished physical activity levels, reaffirming the significant links between depression and physical activity [40-42].

Our findings should be interpreted within the context of several limitations. First, our cohort, which consists predominantly of females with a history of depression and is based in Europe, may limit our generalizability to more diverse or non-depressed populations. Second, over half of our data were collected during the COVID-19 pandemic. Although we employed a covariate to adjust for the effects of national lockdowns in the mediation analysis, varying COVID-19 restrictions over time and across countries make it difficult to fully remove their influence (especially for physical activity) [54]. Therefore, our findings need to be validated in the post-COVID datasets. Third, the weather data used in this study were retrospectively obtained based on the city information provided at the time of PHQ-8 submission. This approach only allowed us to capture the weather conditions on the day of submission, lacking dynamic information during the period before the PHQ-8 assessment. Future digital studies might benefit from concurrently collecting passive data and weather information, providing a richer and more dynamic context for analysis. Fourth, the manually specified thresholds were used in our subgroup analyses. The most appropriate thresholds for subgroup analysis would necessitate further investigations. Fifth, the RADAR-MDD study's use of an open enrollment strategy [48, 72] has resulted in site-specific variations in age, depression severity, and weather conditions (Table 1 and Figure 2). Although we accounted for the site as a covariate, the influence of these site-specific differences on our findings needs further investigation.

In conclusion, this study provides valuable insights into the interplay between weather conditions, physical activity, and depression severity. It highlights the need for personalized approaches in managing depression and underscores the potential of leveraging environmental factors in treatment strategies. Future research should continue to explore these relationships in more diverse populations.

## Data Availability

The processed and anonymized data used for the present study can be made available through reasonable requests to the RADAR-CNS consortium, but the raw passive data and demographics cannot be made available due to participant safety and data privacy issues. Please email the corresponding author for details.

## Code Availability

The code for data analyses will be made available by the corresponding author upon reasonable request.

## Acknowledgments

The Remote Assessment of Disease and Relapse–Central Nervous System (RADARCNS) project has received funding from the Innovative Medicines Initiative (IMI) 2 Joint Undertaking under grant agreement No 115902. This Joint Undertaking receives support from the European Union's Horizon 2020 Research and Innovation Program and the European Federation of Pharmaceutical Industries and Associations (EFPIA). This communication reflects the views of the RADAR-CNS consortium and neither IMI nor the European Union and EFPIA are liable for any use that may be made of the information contained herein. The funding bodies have not been involved in the design of the study, the collection or analysis of data, or the interpretation of data. This study represents independent research partly funded by the National Institute for Health Research (NIHR) Maudsley Biomedical Research Centre at South London, and Maudsley NHS Foundation Trust and King's College London. The views expressed are those of the author(s) and not necessarily those of the NHS, the NIHR, or the Department of Health and Social Care. We thank all the members of the RADAR-CNS patient advisory board for their contribution to the device selection procedures, and their invaluable advice throughout the study protocol design. This research was reviewed by a team with experience of mental health problems and their careers, who have been specially trained to advise on research proposals and documentation through Feasibility and Acceptability Support Team for Researchers (FAST-R), a free, confidential service in England provided by the NIHR Maudsley Biomedical Research Centre via King's College London and South London and Maudsley NHS Foundation Trust. We thank all GLAD Study volunteers for their participation, and gratefully acknowledge the NIHR BioResource, NIHR BioResource centers, NHS Trusts and staff for their contribution. We also acknowledge NIHR BRC, King's College London, South London and Maudsley NHS Trust and King's Health Partners. We thank the NIHR, NHS Blood and Transplant, and Health Data Research UK as part of the Digital Innovation Hub Program. Participants in the CIBER site came from the following four clinical communities in Spain: Parc Sanitari Sant Joan de Déu Network services, Institut Català de la Salut, Institut Pere Mata, and Hospital Clínico San Carlos. Participant recruitment in Amsterdam was partially accomplished through Hersenonderzoek.nl (www.hersenonderzoek.nl), a Dutch online registry that facilitates participant recruitment for neuroscience studies. Hersenonderzoek.nl is funded by ZonMwMemorabel (project no 73305095003), a project in the context of the Dutch Deltaplan Dementie, Gieskes-Strijbis Foundation, the Alzheimer's Society in the Netherlands and Brain Foundation Netherlands.

R.J.B.D. is supported by the following: (1) National Institute for Health and Care Research (NIHR) Biomedical Research Centre (BRC) at South London and Maudsley National Health Service (NHS) Foundation Trust and King's College London; (2) Health Data Research UK, which is funded by the UK Medical Research Council (MRC), Engineering and Physical Sciences Research Council, Economic and Social Research Council, Department of Health and Social Care (England), Chief Scientist Office of the Scottish Government Health and Social Care Directorates, Health and Social Care Research and Development Division (Welsh Government), Public Health Agency (Northern Ireland), British Heart Foundation, and Wellcome Trust; (3) the BigData@Heart Consortium, funded by the Innovative Medicines Initiative 2 Joint Undertaking (which receives support from

## Author Contributions

Y.Z., A.P., A.A.F., and R.J.B.D. contributed to the design of this study. Y.Z. and A.P. designed the data analysis plan and drafted the manuscript. Y.Z. performed data analysis. M.H. is the principal investigator for the RADAR-MDD study. A.A.F., Y.R., P.C., Z.R., C.S., and R.J.B.D., have contributed to the development of the RADAR-base platform for data collection. F.M., C.O., F.L., S. Siddi, S. Simblett, J.M.H., B.W.J.H.P., and M.H. contributed to participant recruitment and data collection. A.A.F., V.A.N., T.W., M.H., and R.J.B.D. contributed to the administrative, technical, and clinical support of the study. All authors were involved in reviewing and providing comments on the manuscript.

## Competing Interests

S.V. is an employee of Janssen Research and Development LLC. V.A.N. was an employee of Janssen Research and Development LLC during the RADAR-MDD study. At the time of manuscript submission, A.P. is an employee of Boehringer Ingelheim Pharmaceuticals, Inc., Ridgefield, CT, USA. M.H. is the principal investigator of the Remote Assessment of Disease and Relapse–Central Nervous System project, a public-private precompetitive consortium that receives funding from Janssen, UCB, Lundbeck, MSD, and Biogen. All other authors declare no competing interests.

# Tables

Table 1. A summary of characteristics of participants in the selected cohort in this study, with comparisons across three study sites using Kruskal-Wallis tests.

| Characteristics | Overall | CIBER (Spain) | KCL (United Kingdoms) | VUMC (Netherlands) | p value |
|---|---|---|---|---|---|
| **Number of participants** | 428 | 90 | 251 | 87 | |
| **All PHQ8, median [IQR]** | 9.5 [6.0,13.7] | 13.4 [8.6,17.5] | 8.9 [5.5,12.5] | 9.1 [6.1,11.9] | <0.001 |
| **Baseline PHQ8, median [IQR]** | 10.0 [6.0,15.0] | 14.5 [10.0,18.8] | 9.0 [6.0,13.0] | 8.0 [6.0,13.0] | <0.001 |
| **Age, median [IQR]** | 50.0 [32.0,60.0] | 55.0 [45.5,61.0] | 47.0 [31.0,59.0] | 40.0 [26.5,58.5] | <0.001 |
| **Years in education, median [IQR]** | 16.0 [13.0,19.0] | 11.0 [9.0,16.0] | 17.0 [14.0,19.0] | 17.0 [14.0,20.5] | <0.001 |
| **Female, n (%)** | 331 (77.3) | 65 (72.2) | 195 (77.7) | 71 (81.6) | 0.322 |
| **Employed, n (%)** | 186 (43.5) | 22 (24.4) | 136 (54.2) | 28 (32.2) | <0.001 |
| **Has children, n (%)** | 211 (49.3) | 69 (76.7) | 113 (45.0) | 29 (33.3) | <0.001 |
| **Married Status, n (%)** | | | | | 0.003 |
| Single | 226 (52.8) | 36 (40.0) | 133 (53.0) | 57 (65.5) | |
| Married | 202 (47.2) | 54 (60.0) | 118 (47.0) | 30 (34.5) | |
| **Annual income (£/€), n (%)** | | | | | <0.001 |
| < 15,000 | 101 (23.6) | 28 (31.1) | 51 (20.3) | 22 (25.3) | |
| 15,000-55,000 | 246 (57.5) | 57 (63.3) | 150 (59.8) | 39 (44.8) | |
| > 55,000 | 70 (16.4) | 5 (5.6) | 50 (19.9) | 15 (17.2) | |

Table 2. Outcomes of mediation models assessing the effect of weather conditions on physical activity via depression severity (PHQ-8) across the entire cohort and subgroups (Positive Response, Negative Response, and Unaffected; see Methods). Significance Levels: *p<0.05, **p<0.01, ***p<0.001.

|  | Temperature (Celsius) | Daylength (hours) | Humidity (%) | Cloudiness (%) | Pressure (hPa) | Wind Speed (meters/sec) |
|---|---|---|---|---|---|---|
| **Entire Cohort** | | | | | | |
| Total Effects (c) | 46.09 *** | 65.6 *** | -7.25 ** | -2.71 * | -1.68 | -0.69 |
| Direct Effect (c') | 44.08 *** | 61.91 *** | -6.53 ** | -2.47 * | -2.65 | 1.96 |
| Indirect Effect (a*b) | 2.01 *** | 3.69 ** | -0.73 * | -0.24 | 0.97 ** | -2.66 |
| a path | -0.02 ** | -0.04 ** | 0.01 * | 0.003 | -0.01 ** | 0.03 |
| b path | -87.84 *** | -89.01 *** | -90.22 *** | -90.36 *** | -91.08 *** | -90.8 *** |
| **Positive Response** | | | | | | |
| Total Effects (c) | 12.1 | 22.14 | -22.78 *** | -8.61 *** | 4.98 | -57.58 |
| Direct Effect (c') | 26.23 * | 64.61 ** | -12.2 * | -5.8 ** | 14.15 | -37.56 |
| Indirect Effect (a*b) | -14.13 *** | -42.47 *** | -10.58 *** | -2.81 *** | -9.17 *** | -20.03 * |
| a path | 0.21 *** | 0.39 *** | 0.09 *** | 0.04 *** | 0.08 *** | 0.47 *** |
| b path | -69.44 *** | -108.95 *** | -112.22 *** | -75.74 *** | -119.04 *** | -43.47 * |
| **Negative Response** | | | | | | |
| Total Effects (c) | 65.54 *** | 131.67 *** | -2.63 | 1.31 | 12.17 * | 107.96 *** |
| Direct Effect (c') | 46.17 *** | 94.85 *** | -8.16 | -1.52 | 2.73 | 85.29 ** |
| Indirect Effect (a*b) | 19.37 *** | 36.82 *** | 5.53 *** | 2.83 *** | 9.44 *** | 22.66 *** |
| a path | -0.19 *** | -0.42 *** | -0.1 *** | -0.04 *** | -0.09 *** | -0.39 *** |
| b path | -102.51 *** | -86.75 *** | -60.17 *** | -67.93 *** | -106.49 *** | -57.27 ** |
| **Unaffected** | | | | | | |
| Total Effects (c) | 47.04 *** | 44.08 *** | -1.18 | -1.6 | -7.06 * | -10.37 |
| Direct Effect (c') | 47.1 *** | 43.53 *** | -0.88 | -1.39 | -7.4 * | -10.0 |
| Indirect Effect (a*b) | -0.06 | 0.56 | -0.31 | -0.22 | 0.34 | -0.37 |
| a path | 0.001 | -0.01 | 0.003 | 0.002 | -0.004 | 0.01 |
| b path | -80.24 *** | -75.38 *** | -87.6 *** | -103.08 *** | -81.73 *** | -111.08 *** |

# Figures

Figure 1. Schematic representation of unmediated and mediated models along with the implementations of mediated pathway analysis in the present study. The mediation model decomposes the total exposure-outcome effect (c path) into a direct effect (c' path) and an indirect effect via a mediator, where the indirect effect is calculated by multiplying the effect of exposure on the mediator (a path) by the effect of the mediator on the outcome (b path). In this paper, each weather variable serves as the exposure, with one model (Model 1) considering depression symptom severity (PHQ-8) as the mediator and physical activity (Step Count) as the outcome, and another (Model 2) reversing these roles.

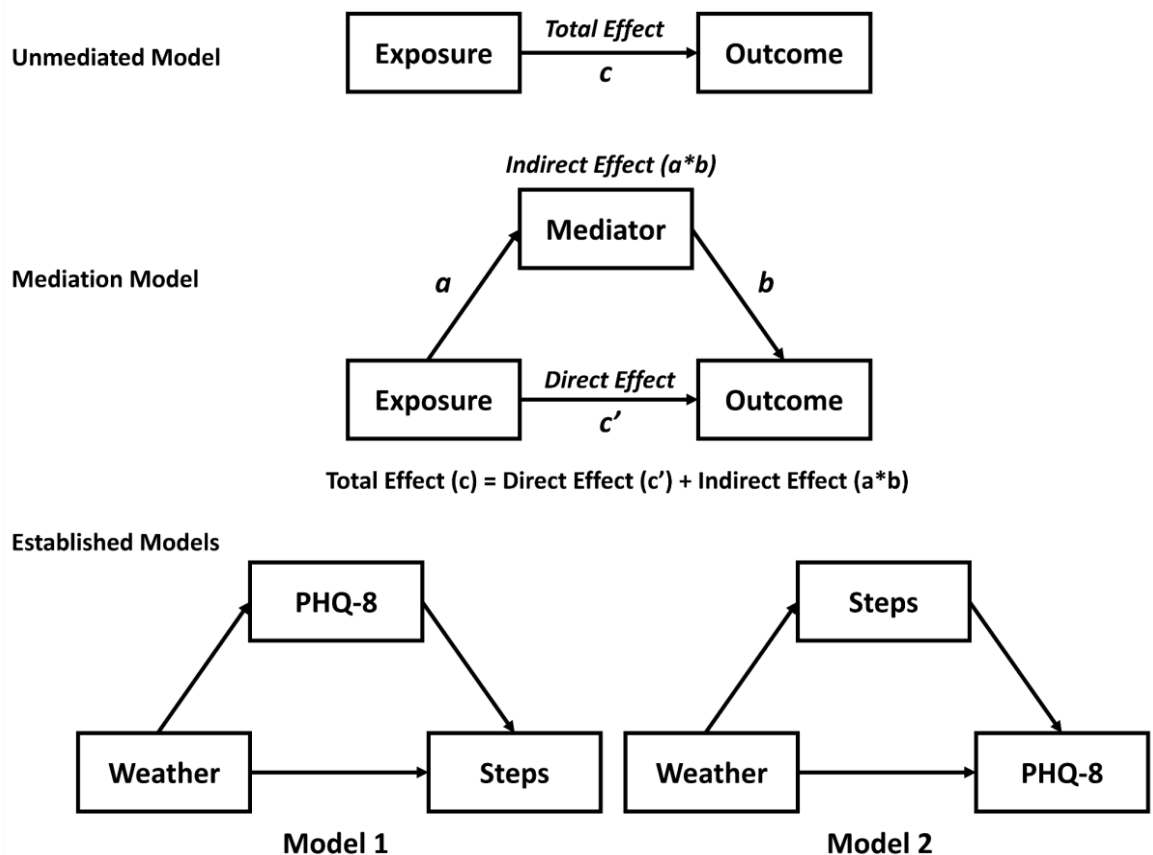

Figure 2. Seasonal variations in weather conditions across three study sites. Weather conditions collected in this study are (a) ambient temperature (in Celsius), (b) day length (the time between sunrise and sunset, measured in hours), (c) humidity (%), (d) cloudiness (%), atmospheric pressure (in hPa), and (e) wind speed (in meters/sec).

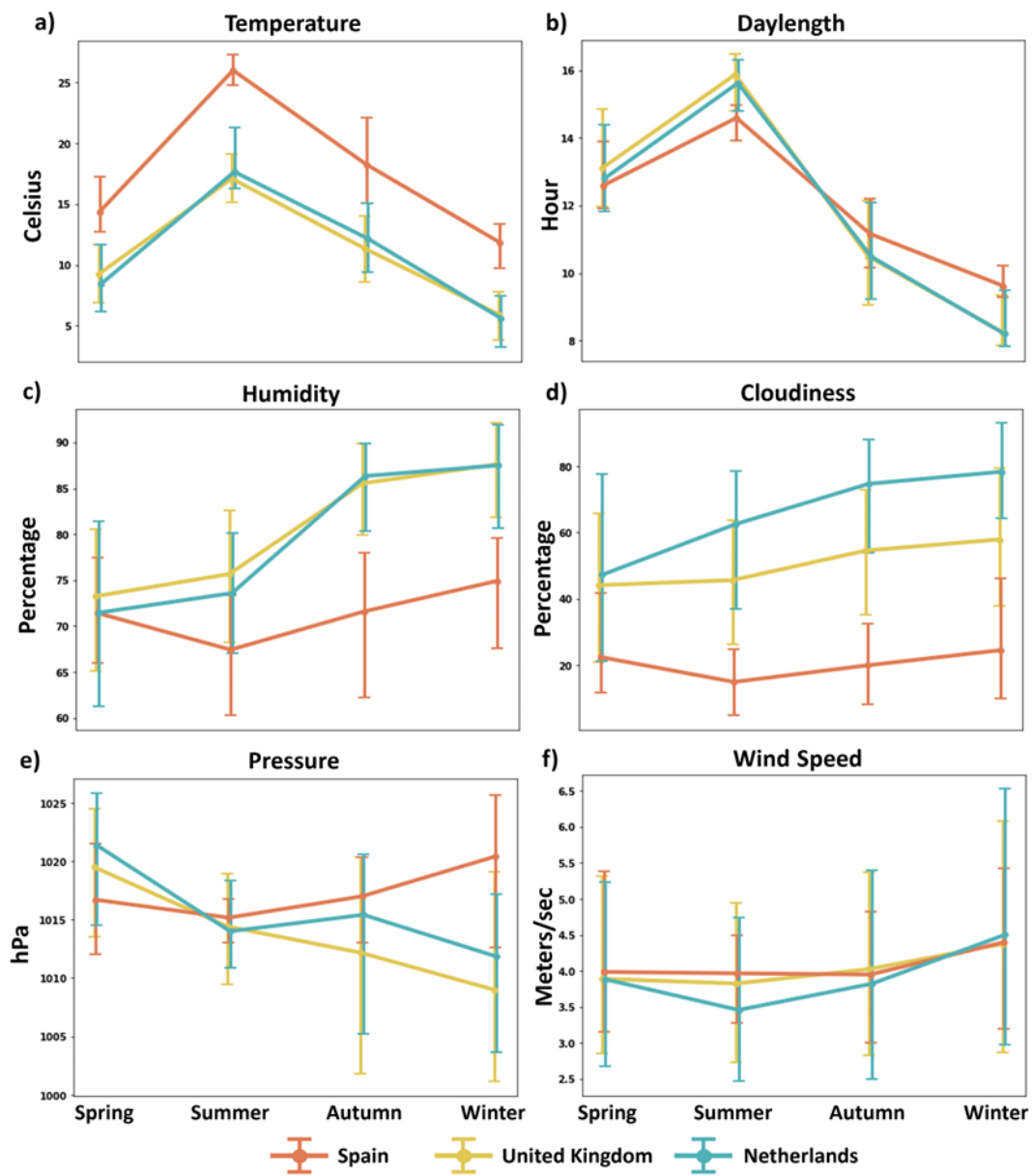

Figure 3. Four distinct patterns of seasonal variations in depression symptom severity (mean-centered PHQ-8 scores; depicted in purple) and corresponding changes in physical activity levels (mean-centered daily steps; depicted in green) within the whole cohort (a-d). Age and baseline PHQ-8 scores are significantly (Kruskal-Wallis tests) different across the four clusters (e-f). The comparisons of other socio-demographics across clusters are shown in Supplementary Table 2.

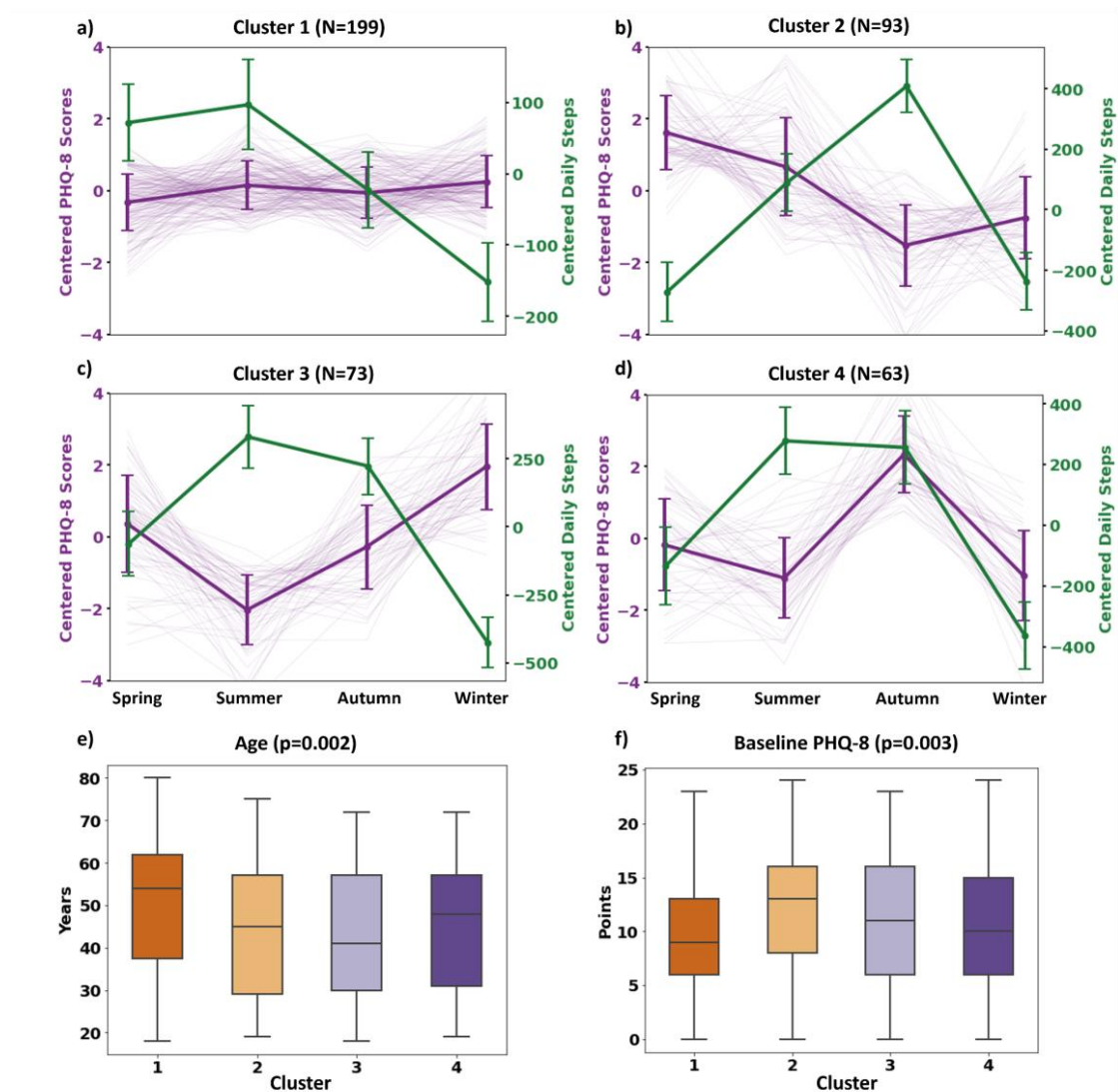

Figure 4. The path diagrams and effects of mediation models for Positive and Negative Response subgroups for temperature and day length. a) Positive Response subgroup for temperature, b) Negative Response subgroup for temperature, c) Positive Response subgroup for day length, and d) Negative Response subgroup for day length. The subgroups were assigned based on Spearman Coefficients between depression severity and weather conditions (see Methods). The orange indicates the positive effect while blue represents the negative effect. The significance levels: * p<0.05, ** p<0.01, *** p<0.001.

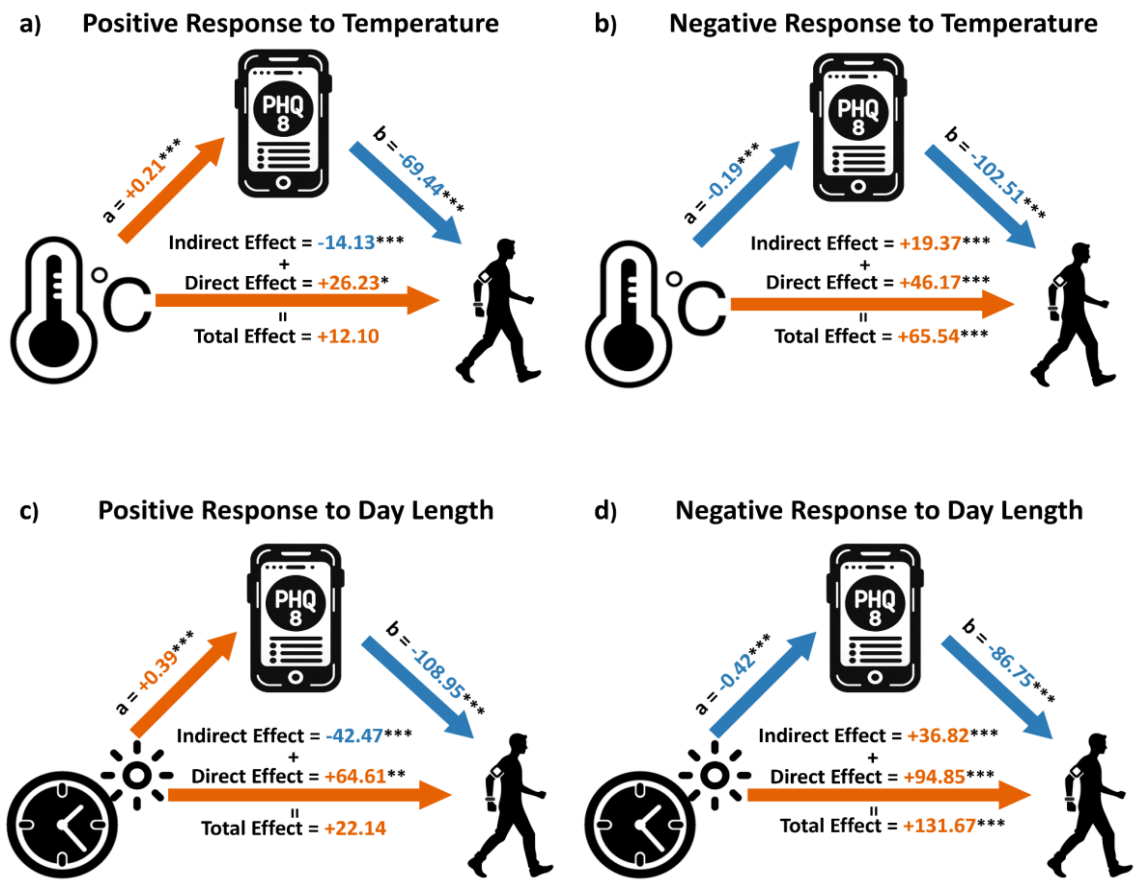